# Geometrical Volume Effects in the Computation of the Slope of the Isgur-Wise Function

*UKQCD Collaboration*

**L. Lellouch[1], J. Nieves, C.T. Sachrajda, N. Stella, H. Wittig**

Physics Department, The University, Southampton SO17 1BJ, UK

**G. Martinelli**

Dipartimento di Fisica, Università di Roma *La Sapienza*, 00185 Roma, Italy.

**D.G. Richards**

Department of Physics, The University of Edinburgh, Edinburgh EH9 3JZ, Scotland

**Abstract**

We use a method recently suggested for evaluating the slope of the Isgur-Wise function, at the zero-recoil point, on the lattice. The computations are performed in the quenched approximation to lattice QCD, on a $24^3 \times 48$ lattice at $\beta = 6.2$, using an $O(a)$-improved action for the fermions. We have found unexpectedly large finite-volume effects in such a calculation. These volume corrections turned out to be purely geometrical and independent of the dynamics of the system. After the study of these effects on a smaller volume and for different quark masses, we give approximate expressions that account for them. Using these approximations we find $\xi'(1) = -1.7\ {}^{+2}_{-2}$ and $\xi'(1) = -1.4\ {}^{+2}_{-1}$ for the slope of the Isgur-Wise function, for two mesons composed of a heavy quark slightly heavier and lighter, respectively, than the charm quark, and in both cases, a light antiquark whose mass is about that of the strange quark.

---

[1]Present address: Centre de Physique Theorique, CNRS Luminy, Case 907, F-13288 Marseille Cedex 9, France.

# 1 Introduction

In recent years there has been an increased theoretical and experimental interest in the study of decays of hadrons containing a $b$ quark. Due to the appearance of a spin-flavour symmetry in the dynamics of heavy quarks ([1]), systems involving a heavy quark are in some ways simpler than those involving only light quarks. The study of these systems has several objectives, including the extraction of the elements $V_{cb}$ and $V_{ub}$ of the Cabibbo-Kobayashi-Maskawa matrix, a determination of the applicability of the "Heavy Quark Effective Theory" (HQET) predictions for the physics of $b$ and $c$ quarks, and derivations of bounds for the effects of physics beyond the standard model.

Recently the ARGUS [2] and CLEO [3] collaborations have measured the semileptonic decay $B \to D^* l \bar{\nu}$, and have presented results for $|V_{cb}|\bar{\xi}(\omega)$, where $\bar{\xi}(\omega)$ is the Isgur-Wise function, $\xi(\omega)$, up to short-distance and heavy-quark mass corrections [4] and $\omega$ is the dot product of the four-velocities of the $B$ and $D^*$ mesons. The Isgur-Wise function is the unique form-factor for all semi-leptonic $B \to D$-decays in the limit of infinite $b$ and $c$ quark masses. The extraction of the $|V_{cb}|$ matrix element requires a precise knowledge of the Isgur-Wise function in a region of values of $\omega$ close to the zero-recoil point (i.e. $\omega = 1$; the Isgur-Wise function is normalised such that $\xi(1) = 1$).

Lattice computations allow for the determination of the Isgur-Wise function for discrete values of $\omega > 1$ ([5] - [6]) and these can be used to reconstruct the Isgur-Wise function near to the zero-recoil point, and in particular to estimate its derivatives at $\omega = 1$. Although much of the systematics is understood in this kind of calculation, [7], there are still some problems which lead to uncertainties in the determination of $\xi(\omega)$ near to the zero-recoil point (extrapolation of the results from $\omega > 1$ to $\omega = 1$, determination of the possible $\omega$-dependence of the systematic effects which may distort this extrapolation, etc...) and could lead to uncertainties in the extraction of $|V_{cb}|$. For this reason, the authors of [8] suggested a new method, in which the slope and higher derivatives of the Isgur-Wise function at $\omega = 1$ are computed directly on the lattice, hence avoiding any kind of extrapolation (which necessarily relies on a certain parametrization of the $\omega$-dependence of Isgur-Wise function) from $\omega > 1$ to the zero-recoil point. The results obtained by using this new method would complement those obtained by using the traditional one ([5] - [6]) and would help to unravel possible sources of systematic errors in lattice calculations.

In future it is likely that it will also be possible to study the Isgur-Wise function near the zero-recoil point by simulations using the HQET. Results from an early study can be found in ref. [9].

In [8] it was proposed to study correlators of the type:

$$C_2^{(2)}(t) = -\frac{1}{2} \sum_{\vec{x}} x_3^2 \langle 0|J_P(x) J_P^\dagger(0)|0\rangle, \tag{1}$$

$$C_3^{(2)}(t, t_y) = -\frac{1}{2} \sum_{\vec{x},\vec{y}} x_3^2 \langle 0|J_P(y) V^4(x) J_P^\dagger(0)|0\rangle, \tag{2}$$

in addition to the usual two- and three-point correlators, ($C_2^{(0)}(t)$ and $C_3^{(0)}(t)$) given by:

$$C_2^{(0)}(t) = \sum_{\vec{x}} \langle 0|J_P(x) J_P^\dagger(0)|0\rangle, \tag{3}$$

$$C_3^{(0)}(t, t_y) = \sum_{\vec{x},\vec{y}} \langle 0|J_P(y) V^4(x) J_P^\dagger(0)|0\rangle, \tag{4}$$

where $J_P(J_P^\dagger)$ is an interpolating operator which can annihilate (create) the pseudoscalar meson $P$, and $V^4$ represent the fourth component of the vector current $\bar{Q}\gamma^\mu Q$. In [8] it was shown that, from the measurement of these correlators, the slope of the Isgur-Wise function at the zero-recoil point could be extracted. Actually, if local interpolating operators are used and in the infinite-volume limit, it can be shown that:

$$\frac{1}{2m^2}\left(\frac{1}{2} + \frac{d\bar{\xi}}{d\omega}(1)\right) = R_3^{(2)} - R_2^{(2)}, \tag{5}$$

where $\bar{\xi}(\omega)$, as we mentioned above, is the Isgur-Wise function up to short-distance and heavy-quark mass corrections and the ratios $R_{3,2}^{(2)}$ are defined by:

$$R_{2,3}^{(2)} \equiv \frac{C_{2,3}^{(2)}}{C_{2,3}^{(0)}}. \tag{6}$$

The second derivative of the Isgur-Wise function would be determined by a similar set of correlators to those of eqs.(1-2) in which $x_3^2$ has been replaced by $x_3^4$, and in principle, higher order derivatives could be determined alike [8].

Below we present the results of an exploratory study to check the feasibility of the method proposed in [8]. The main result of this paper is that in the evaluation of correlation functions of the form in eqs.(1) and (2) there are large finite-volume effects. Although these effects, for a fixed meson mass and time, decrease exponentially with the length of the lattice in the spatial directions (with a rate given, to a good approximation, by $(m/t)^{\frac{1}{2}}$), they are large on currently available lattices. The origin of these effects is purely geometrical and understandable, and they do not depend on the dynamics of the system. The ordinary two- and three-point correlators (without the insertion of the $x_3^2, x_3^4, ...$ terms) used in lattice calculations are free of this new source of systematic errors. The problem arises because when the terms $x_3^2, x_3^4, ...$ are included in the definition of the correlators, the lattice theory becomes sensitive to momentum contributions in addition to the unique momentum mode which determines the ordinary two and three point correlators[2]. The behaviour of these geometrical volume effects, and in particular, the dependence on the mass and time of evolution, is different from that in usual correlation functions. Similar effects will be present in any direct computation of a derivative of a matrix element with respect to an external momentum. Since the origin of these effects is purely kinematical, they can be controlled, and below we present a set of approximations which will allow us to extract the slope of the Isgur-Wise function at zero-recoil in simulations on finite volumes. Using these approximations we find $\xi'(1) = -1.7 \, {}^{+2}_{-2}$ and $\xi'(1) = -1.4 \, {}^{+2}_{-1}$, for two mesons composed of a heavy quark slightly heavier and lighter, respectively, than the charm quark, and in both cases, a light antiquark whose mass is about that of the strange quark. These values agree well with those obtained in the lattice computation of ref. [7]. In this reference, for the same heavy and light quark masses and using the traditional method of extrapolating from $\omega > 1$ to the zero-recoil point, were found the following values for the slope of the Isgur-Wise function: $\xi'(1) = -1.4 \, {}^{+4}_{-3}$ and $\xi'(1) = -1.4 \, {}^{+2}_{-2}$ respectively.

The remainder of this paper is organized as follows. In Section 2 we will present and study the origin of these new volume effects in the simplest case: two-point function correlators with local interpolating operators $J_P$. The results of this section will permit us to study in Section 3 the case of extended interpolating operators and the computation of the slope of the Isgur-Wise function at zero-recoil. Finally, in Section 4 we will present our conclusions.

## 2  Geometrical Finite-Volume Effects

In this section we demonstrate the existence of finite-volume effects in correlation functions of the form in eq. (1), and show that these effects are large on currently available lattices. Consider the evaluation of such a correlation function on a $N^3 \times N_T$ lattice, of lengths $T$ and $L$ in the temporal and spatial directions respectively, and with lattice spacing $a$ ($L = Na, T = N_T a$):

$$C_2^{(2)}(t) = -\frac{1}{2} \sum_{\vec{x}} x_3^2 \langle 0 | J_P(x) J_P^\dagger(0) | 0 \rangle, \tag{7}$$

$$= -\frac{1}{2} \sum_A \sum_{x_3=a(-\frac{N}{2}+1)}^{a\frac{N}{2}} x_3^2 \frac{1}{N} \sum_{n=-\frac{N}{2}+1}^{\frac{N}{2}} Z_A^2(n^2 p_{min}^2) \frac{e^{-(\sqrt{m_A^2+p_{min}^2 n^2})\,t}}{2\sqrt{m_A^2+p_{min}^2 n^2}} e^{i\,np_{min}x_3}, \tag{8}$$

where $p_{min} = 2\pi/L$, $Z_A(\vec{p}^2) = |\langle 0 | J_P(0) | \vec{p}, A \rangle|$ and the sum $\sum_A$ is over all the excited states

---

[2]This momentum mode coincides with the external momentum provided in the definition of the correlators.

$A$ that can contribute to the correlation function. $m_A$ is the mass of state $A$[3]. In general $Z(\vec{p}^{\,2})$ depends on the momentum $|\vec{p}|$ if extended interpolating operators $J_P$ are used.

In the infinite-volume limit $L, T \to \infty$, and with $t \gg 1$ so that only the ground state contributes significantly[4], the correlation function becomes [8]:

$$C_2^{(2)}(t) \approx C_2^{(0)}(t) \times \left(-\frac{1}{2m^2} - \frac{t}{2m} + 2R_Z^{(2)}\right), \tag{9}$$

where

$$C_2^{(0)}(t) = \frac{Z^2(0)}{2m} e^{-mt}, \tag{10}$$

$$R_Z^{(2)} = \frac{Z'(0)}{Z(0)}, \tag{11}$$

$$Z'(\vec{p}^{\,2}) = \frac{dZ(\vec{p}^{\,2})}{d\vec{p}^{\,2}}. \tag{12}$$

We now study the finite-volume corrections to eq. (9). For purposes of illustration let us take the interpolating operators $J_P(x)$ to be local ones, so that the corresponding wavefunction factor $Z(\vec{p}^{\,2})$, in eq. (8), is independent of momentum, and can be taken out of the sum over $n$, and also $R_Z^{(2)} = 0$. Having determined the mass of the ground state from the behaviour of the correlation function $C_2^{(0)}(t)$ with time (see eq. (10) ), the ratio of $C_2^{(2)}(t)$ and $C_2^{(0)}(t)$ can be calculated exactly by computing the discrete sum in eq. (8):

$$R_2^{(2)}(t) \equiv \frac{C_2^{(2)}(t)}{C_2^{(0)}(t)}, \tag{13}$$

$$= -\frac{1}{2} \sum_{x_3 = a(-\frac{N}{2}+1)}^{a\frac{N}{2}} x_3^2 \frac{1}{N} \sum_{n=-\frac{N}{2}+1}^{\frac{N}{2}} \frac{e^{-(\sqrt{m^2 + p_{min}^2 n^2} - m) t}}{\sqrt{1 + p_{min}^2 n^2/m^2}} e^{i n p_{min} x_3}. \tag{14}$$

Assuming that:

i) The lattice spacing in the spatial directions is zero, but that $L$ is finite, so that the sum over $x_3$ can be replaced by an integral, i.e.

$$\frac{1}{N} \sum_{x_3 = a(-\frac{N}{2}+1)}^{a\frac{N}{2}} \to \frac{1}{L} \int_{-\frac{L}{2}}^{\frac{L}{2}} dx_3, \tag{15}$$

which implies that the sum over $n$ in eq. (14) must be extended between $-\infty$ to $+\infty$.

---

[3]There may also be contributions from multiparticle states, but we will shortly restrict the discussion to the ground state only.

[4]In the following, except where explicitly stated, we will neglect the contributions from the excited states.

ii) $L \gg 1$ and therefore we can perform an asymptotic expansion in powers of $1/L$.

the result of eq. (14) can be approximated by:

$$R_2^{(2)}(t) \approx \left(-\frac{1}{2m^2} - \frac{t}{2m}\right) \times \left(1 - \frac{L}{(\frac{3}{m^2} + \frac{t^2}{1+mt})^{\frac{1}{2}}} \frac{1}{\sinh(L/(\frac{3}{m^2} + \frac{t^2}{1+mt})^{\frac{1}{2}})}\right). \qquad (16)$$

In general eq. (16) is a good approximation to eq. (14) when both $\pi/mL$ and $\pi^2 t/mL^2$ are small quantities; the higher the mass $m$, and the smaller the time $t$, the better is the approximation. In Figs.1 and 2 we test the validity of eq. (16) by plotting $R_2^{(2)}(t)$ as a function of the time, t, for different volumes and masses of the scalar particle. The plots correspond to masses given by $ma = 0.3$ and $0.6$, and volumes $V = 24^3$ and $36^3$. In each case the dashed line represents $R_2^{(2)}(t)$ computed numerically from the exact expression given in eq. (14), the solid line corresponds to the infinite-volume result, $R_2^{(2)}(t) = (-1/2m^2 - t/2m)$, and the squares stand for the approximation of eq. (16) to $R_2^{(2)}$. The first observation which can be made from these plots is that there is a considerable difference between infinite-(solid line) and finite-(dashed line) volume predictions. A second observation is that eq. (16) is in most cases a good approximation to the finite-volume prediction of eq. (14), i.e. for a wide range of volumes, masses and times, the squares follow the dashed line.

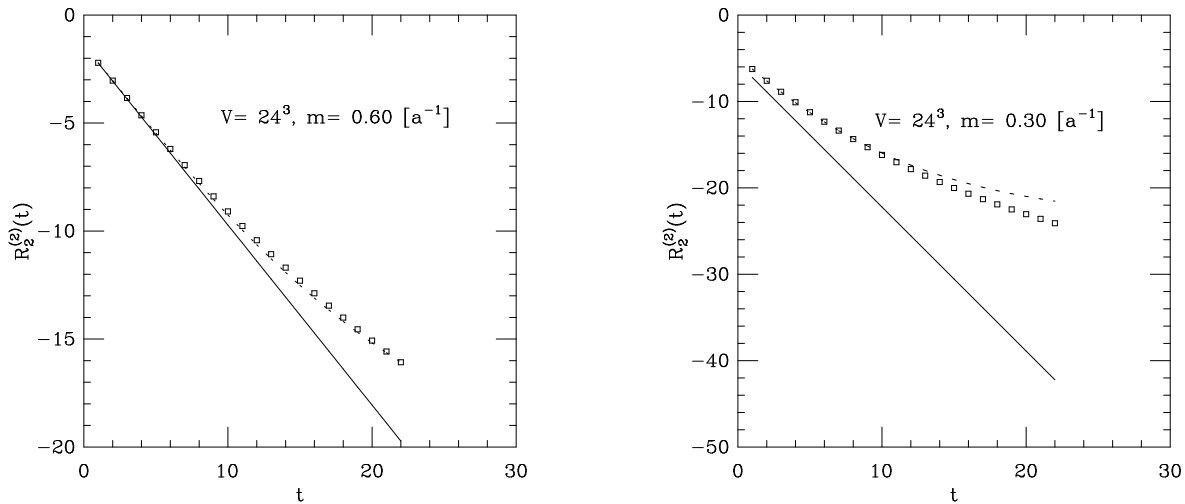

Figure 1: Comparison between different approximations to $R_2^{(2)}(t)$ (see text). a) m=0.6; b) m=0.3. In both cases $V = 24^3$.

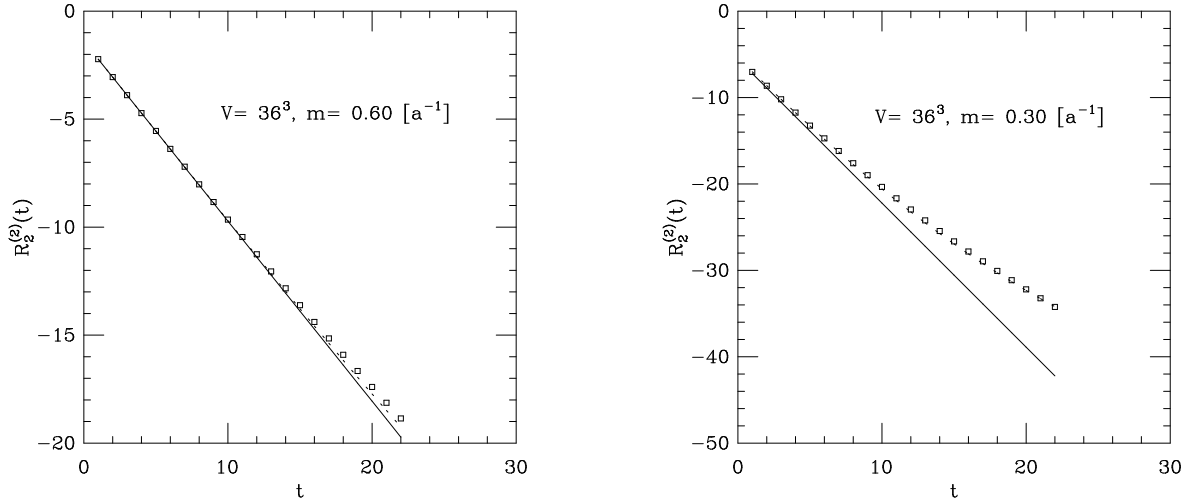

Figure 2: Comparison between different approximations to $R_2^{(2)}(t)$ (see text). a) m=0.6 ; b) m=0.3. In both cases $V = 36^3$ .

From eq. (16) we see that the finite-volume effects decrease exponentially with $L$ and, for sufficiently large times, they are given by:

$$\frac{R_2^{(2)}(t)}{R_2^{(2)}(t, L = \infty)} = 1 - 2L\sqrt{\frac{m}{t}} e^{-L\sqrt{\frac{m}{t}}}, \qquad (17)$$

where the dependence on $L$, $m$ and $t$ is explicitly shown.

The origin of these geometrical volume corrections is due to the fact that more than one mode contributes to the sum over $n$ in eq. (8). By contrast, in the usual two-point functions, $C_2^{(0)}(t)$, only the term with $n = 0$ contributes in the sum, and thus this type of geometrical finite-volume effect is not present. However, both $C_2^{(0)}(t)$ and $C_2^{(2)}(t)$ suffer from dynamical volume effects. These effects are either due to self interactions of the particle which is being studied with its infinite number of mirror copies (in simulations with periodic boundary conditions) [10], or due to the modification of the hadronic wave function in situations for which the physical volume of the lattice is comparable to the size of the hadron [11]-[12]. For small lattices, the latter type of corrections is the dominant effect, whereas for physical volumes of the order of 1.5 fm the predominant effect is due to the interaction of the particle with its mirror copies ([10],[12]). However, all of these corrections are small for the masses and volumes considered in this paper, and in what follows they will be ignored.

As was mentioned in the introduction, the slope of the Isgur-Wise function in the infinite-volume limit can be extracted from the correlators $C_2^{(2)}(t)$ and $C_3^{(2)}(t)$. For finite volumes such a determination contains similar corrections to the ones studied here for $C_2^{(2)}(t)$, corrections which decrease exponentially with the length of the lattice. The slope of the Isgur-Wise function at $\omega = 1$ can also be estimated by computing the finite difference $\xi(\omega_{min}) - \xi(1)$, where $\omega_{min}$ is the discrete value of $\omega$ nearest to $\omega = 1$ available on the lattice being used. $\xi'(1)$ derived in such a way suffers from finite-volume corrections of the order of $O(\omega_{min})$ or equivalently of the order of $O(1/L^2)$. These finite-volume corrections, therefore, formally decrease more slowly with the volume of the lattice than those affecting $C_2^{(2)}(t)$ and $C_3^{(2)}(t)$.

In the remainder of this section we present the results from numerical simulations using local interpolating operators. We have measured $R_2^{(2)}(t)$ for a heavy meson with a mass of about 1.6 GeV in a study on a $24^3 \times 48$ lattice at $\beta = 6.2$ and using the $O(a)$-improved quark action ($SW$ or "clover") proposed by Sheikholeslami and Wohlert, ([14, 15])[5]. The meson was composed of a heavy quark with a mass corresponding to $\kappa_h = 0.133$ and a light antiquark with $\kappa_l = 0.14144$ [6]. The mass of the meson was found (from 60 gauge field configurations) to be $ma = 0.598 \,{}^{+3}_{-1}$ [13].

In Figure 3 we compare the prediction of eq. (14) with the data. The solid line corresponds to the infinite-volume prediction $R_2^{(2)}(t) = (-1/2m^2 - t/2m)$, the dashed line to $R_2^{(2)}(t)$ computed using eq. (14) [7] (including volume effects) and finally the squares stand for the measured data (from 10 gauge field configurations) on the lattice. From Figure 3 it is clear that the dashed line fits the data much better than the solid line, demonstrating that the discrepancy between the infinite-volume prediction (9) and the data is largely due to the geometrical finite-volume effects. However, even the dashed line does not reproduce the data perfectly, and we understand the small discrepancy between the two as being due to discretisation errors (i.e. errors due to the finiteness of the lattice spacing)[8]. This will be discussed further below. We have checked that the small discrepancy between the dashed line and the data is not due to the contribution of excited states, since the size of such a contribution would be incompatible with the behaviour of the two-point function $C_2^{(0)}(t)$.

We have also computed the ratio $R_2^{(2)}(t)$ on 36 gauge configurations, for a light meson composed of a degenerate (light) quark – anti-quark pair, in a simulation on a $16^3 \times 32$ lattice at $\beta = 6.0$, again using the $SW$-improved action for the fermions[9]. The mass of the light quark corresponds to $\kappa = 0.1425$ (and that of the corresponding meson was found to be $ma = 0.448\,{}^{+7}_{-4}$); in lattice units the quark mass is about four times smaller than the heavy quark mass in the simulation described above. Thus the discretisation errors should be reduced considerably while the finite-volume effects are going to be even more pronounced, and indeed we now find that this is the case. In Figure 4 we compare the prediction of eq. (14) with the measured data. The meaning of the different curves is the same as in Figure 3. It can be seen from Figure 4 that the agreement between our theoretical prediction and the data is excellent, in spite of the fact that the finite-volume corrections in the ratio $R_2^{(2)}(t)$ are very large for such light mesons.

---

[5] For general details of the simulation, which was performed on the 64-node i860 Meiko Computing Surface at the University of Edinburgh, see references [13] and [16].

[6] In this simulation the hopping parameter of a massless quark is $\kappa_{crit} = 0.14315(2)$, and that of the strange quark is $\kappa_s = 0.1419(1)$ ([13]).

[7] We have appropriately modified eq. (14) in order to include the contribution of the meson propagating backwards in time.

[8] The $O(a)$-improved action which has been used in this work is free of $m_Q a$ discretisation errors. The leading discretisation errors are of order $\alpha m_Q a$ or $(m_Q a)^2$ ([15]) which for a heavy quark may be of a comparable size. $m_Q$ is the mass of the heavy quark.

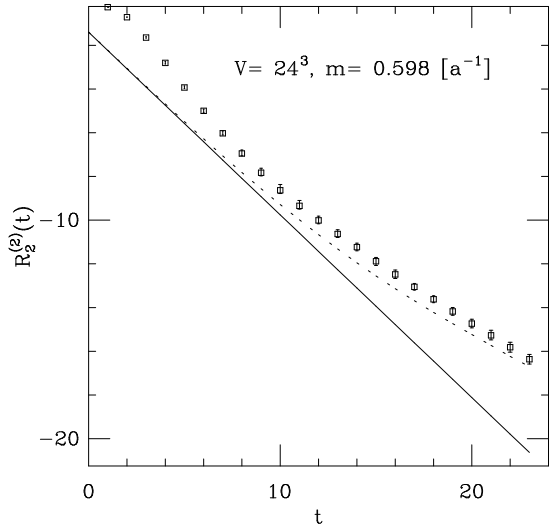

Figure 3: Local interpolating operators: Comparison between the approximation of eq. (14) (dashed line), the infinite-volume prediction (solid line) and the measured data (squares). $\kappa_h = 0.133, \kappa_l = 0.14144, \beta = 6.2$ on $24^3 \times 48$ lattice.

We now try to justify our hypothesis that the small discrepancy between the theoretical prediction (dashed line) and data (squares) in Figure 3 is due to discretisation errors. We have tried to reduce these errors by replacing the continuum energy-momentum dispersion relation in eq. (14) by the lattice dispersion relation of a free boson and the residue factor $1/2E$ by $1/2\sinh(E)$ ([18]). More precisely we have replaced in eq. (14):

$$e^{-(\sqrt{m^2+p_{min}^2 n^2}-m)\,t} \quad \rightarrow \quad e^{-(E^{LA}(m^2,p_{min}^2 n^2)-m)\,t}, \tag{18}$$

---
[9]This simulation was performed on the CRAY-YMP at Cineca in Bologna, and details of the simulation can be found in ref.[17].

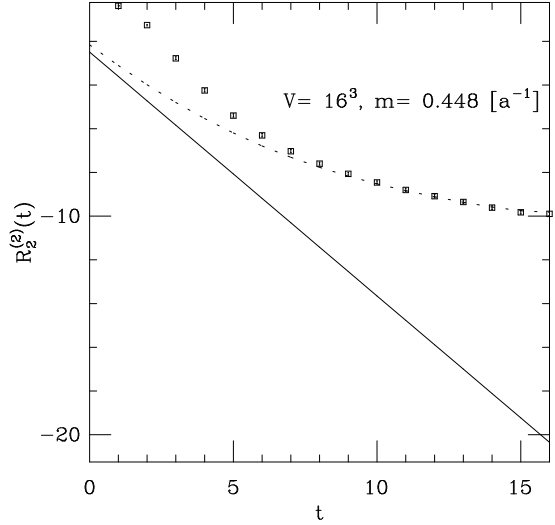

Figure 4: Local interpolating operators: Comparison between the approximation of eq. (14)(dashed line), the infinite-volume prediction (solid line) and the measured data (squares). $\kappa_l = 0.1425, \beta = 6.0$ on $16^3 \times 32$ lattice.

$$\frac{1}{\sqrt{1 + p_{min}^2 n^2/m^2}} \quad \to \quad \frac{\sinh(ma)}{\sinh(E^{LA}(m^2, p_{min}^2 n^2)a)}, \tag{19}$$

where

$$E^{LA}(m^2, q^2) = \frac{2}{a} \text{arc}\sinh\left(\sqrt{\sinh^2(\frac{ma}{2}) + \sin^2(\frac{qa}{2})}\right). \tag{20}$$

Having made these changes, we again compare our prediction, now given by eq. (14) and eqs.(18–20), with the data for the ratio $R_2^{(2)}(t)$ of Figure 3. The result is plotted in Figure 5. The meaning of the different curves is again the same as in Figure 3. The agreement of the predictions with the data is excellent. This fact can be further underlined by noting that if we fit $R_2^{(2)}(t)$ to the theoretical prediction with the mass left as a free parameter we obtain $ma = 0.60(1)$, in good agreement with the value of $0.598\,^{+3}_{-1}$ obtained from $C_2^{(0)}(t)$. The result of such a fit is $ma = 0.63(1)$ when the substitutions of eqs.(18-20) are not taken into account.

In the light-light case of Figure 4, the inclusion of finite-lattice-spacing effects leads to minor changes, and a slightly improved $\chi^2$ in the comparison of the ratio $R_2^{(2)}(t)$ with the theoretical prediction.

In summary, we repeat that, by making a reasonable ansatz for the discretisation errors,

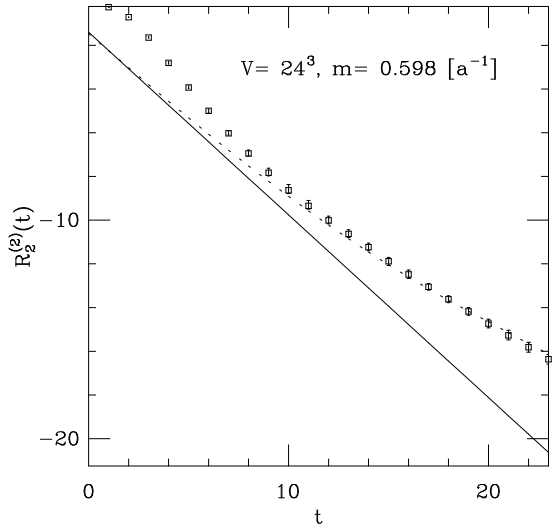

Figure 5: Local interpolating operators: Comparison between the approximation of eq. (14) and eqs. (18 - 20) (dashed line), the infinite-volume and zero lattice spacing prediction (solid line) and the measured data (squares). $\kappa_h = 0.133, \kappa_l = 0.14144, \beta = 6.2$ on $24^3 \times 48$ lattice.

we can account for the small discrepancy between the theoretical finite-volume prediction and the data in Figure 3. Moreover, for light mesons (Figure 4), there is no corresponding discrepancy. These observations, and the fact that the discrepancy cannot be explained by contamination by excited states, leads us confidently to attribute it to the presence of discretisation errors. We stress, however, that the main point of this section was the demonstration that the infinite-volume prediction for $R_2^{(2)}$ has large, controllable, finite-volume effects, as is clear from the difference between the solid and dashed lines in figures 1–5.

## 3   The Slope of the Isgur-Wise function

In this section we will extract the slope of the Isgur-Wise function by computing two- and three-point correlation functions, using extended (or "smeared") interpolating operators in order to enhance the contribution of the ground state. In this exploratory study we restrict the computation to two values of the mass of the heavy quark, and to a single value of the mass of the light quark (as explained below).

Following the ideas of [8], let us consider now[10],

$$C_3^{(2)}(t, t_y) = -\frac{1}{2} \sum_{\vec{x},\vec{y}} x_3^2 \langle 0|J_P(y)V^4(x)J_P^\dagger(0)|0\rangle \tag{21}$$

$$= -\frac{Z_V}{2}\frac{Z(0)}{2m}e^{-m(t_y-t)} \sum_{x_3=a(-\frac{N}{2}+1)}^{a\frac{N}{2}} x_3^2 \frac{1}{N} \sum_{n=-\frac{N}{2}+1}^{\frac{N}{2}} Z(n^2 p_{min}^2) \times M^4(n^2 p_{min}^2)$$

$$\times \frac{e^{-(\sqrt{m^2+p_{min}^2 n^2})\,t}}{2\sqrt{m^2+p_{min}^2 n^2}} \, e^{i n p_{min} x_3}, \tag{22}$$

where $V^4$ represents the fourth component of the vector current $Z_V \bar{Q}\gamma^\mu Q$. $Z_V$ is the renormalisation constant relating the lattice vector current to the physical one. $M^4$ is the fourth component of the matrix element defined by

$$M^\mu(n) \equiv \langle P(\vec{p'})|\bar{Q}\gamma^\mu Q|P(\vec{p})\rangle = (p+p')^\mu \bar{\xi}(\omega), \tag{23}$$

where $\vec{p'} = \vec{0}$, $\vec{p} = (0, 0, np_{min})$ and

$$\bar{\xi}(\omega) = (1 + \beta(\omega) + \gamma(\omega))\xi(\omega), \tag{24}$$

$$\omega = 1 - \frac{q^2}{2m^2}, \tag{25}$$

$$q = p - p'. \tag{26}$$

$\beta(\omega)$ represents the radiative corrections and $\gamma(\omega)$ corresponds to corrections proportional to inverse powers of the heavy-quark mass ($\beta(1) = \gamma(1) = 0$ for the degenerate case we are considering here [1], where by degenerate we mean that the initial and final state mesons have the same masses); in what follows we will neglect the corrections $\gamma(\omega)$, which appear to be very small in the range of masses studied in this work [7]. $\xi(\omega)$ is the Isgur-Wise function and $\omega$ is the dot product of the four velocities of the initial and final state mesons. The Isgur-Wise function is normalised to 1 at the zero-recoil point, i.e. $\xi(1) = 1$. In eq. (22) it has been assumed that $t \gg 1$ and $t_y - t \gg 1$ and that therefore only the ground state contributes to the correlation function.

In the limit $L \to \infty$ the correlation function behaves like[11]:

$$R_3^{(2)}(t) \equiv \frac{C_3^{(2)}(t, t_y)}{C_3^{(0)}(t_y)} \approx \left(-\frac{1}{2m^2} - \frac{t}{2m} + R_Z^{(2)} + R_M^{(2)}\right), \tag{27}$$

where

$$C_3^{(0)}(t_y) = Z_V \frac{Z^2(0)}{4m^2} e^{-mt_y} M^4(0), \tag{28}$$

---

[10]For the present we will ignore possible discretisation errors. We will come back to this point later.

[11]Note that the ratio $R_3^{(2)}(t)$ does not depend on the renormalisation constant $Z_V$.

$$R_M^{(2)} = \frac{M^{4\prime}(0)}{M^4(0)} = \frac{1}{2m^2}(\frac{1}{2} + \frac{d\bar{\xi}}{d\omega}(1)), \tag{29}$$

$$\frac{d\bar{\xi}}{d\omega}(1) = \frac{d\xi}{d\omega}(1) + \frac{d\beta}{d\omega}(1), \tag{30}$$

$$M^{4\prime}(\vec{p}^2) = \frac{dM(\vec{p}^2)}{d\vec{p}^2} \tag{31}$$

and $R_Z^{(2)}$ was defined in eq. (11). The idea of the authors of ref.[8] was to obtain the derivative of the renormalisation-group-invariant Isgur-Wise function $\xi(\omega)$ at the zero-recoil point ($\omega = 1$) simply in terms of the ratios $R_{2,3}^{(2)}$. In the limit $L \to \infty$, the derivative[12] of the Isgur-Wise function at $\omega = 1$ is given by $R_3^{(2)} - R_2^{(2)}$,

$$\frac{1}{2m^2}(\frac{1}{2} + \frac{d\bar{\xi}}{d\omega}(1)) = R_3^{(2)} - R_2^{(2)} + R_Z^{(2)}. \tag{32}$$

As was pointed out in [8], this method allows for a direct determination of the slope of the Isgur-Wise function at zero-recoil, without the need for an extrapolation of the computed values of $\xi(\omega)$ to $\omega = 1$. Moreover the derivatives are computed in terms of ratios of correlation functions, and it is likely that there is some cancellation of systematic errors and statistical fluctuations in these ratios.

We have measured the ratios $R_{2,3}^{(2)}$ using 49 gauge field configurations in our simulation on the $24^3 \times 48$ lattice at $\beta = 6.2$ discussed in the previous section. As mentioned above, the computations were performed using extended interpolating operators. The slope was determined for two values of the mass of the heavy quark (corresponding to $\kappa_h = 0.125, 0.133$, slightly heavier and lighter, respectively, than the charm quark mass, [16]) and for a single value of the light quark mass ($\kappa_l = 0.14144$), which corresponds approximately to the mass of the strange quark [13].

We now discuss, in some detail, the evaluation of the ratios $R_3^{(2)}$ and $R_2^{(2)}$ computed using extended interpolating operators. In particular, the discussion of the geometrical finite-volume effects of the previous section will have to be generalised to the case of extended interpolating operators[13].

---

[12]In order to obtain $\xi'(\omega)$ from the results for $\bar{\xi}'$, it is necessary to evaluate the derivative of $\beta(\omega)$. This is a calculation in continuum perturbation theory and for $m_Q$ about the charm quark mass, ref. [8] found $\beta'(1) = -0.24$.

[13]For $R_3^{(2)}$, even if local interpolating operators are used, eq. (22) can not be computed exactly unless the dependence of the matrix element $M^4(n)$ on $n$ is known.

## 3.1 The ratio $R_2^{(2)}(t)$ computed using Extended Interpolating Operators

When extended interpolating operators are used, the wavefunction factor $Z(\vec{p}^2)$ in eq. (8), depends on $|\vec{p}|$ and therefore can not be taken out of the sum over $n$. Thus, the exact computation of $R_2^{(2)}$ would require the knowledge of the wavefunction factors for all possible momenta of the lattice. However, in the limit $L \gg 1$, the dependence of $Z(n^2 p_{min}^2)$ on momentum can be approximated by:

$$Z(n^2 p_{min}^2) \approx Z(0) \left(1 + (\frac{2\pi}{L})^2 n^2 R_Z^{(2)} + (\frac{2\pi}{L})^4 n^4 R_Z^{(4)}\right), \tag{33}$$

$$R_Z^{(4)} = \frac{1}{2} \frac{Z''(0)}{Z(0)}, \tag{34}$$

$$Z''(\vec{p}^2) = \frac{d^2 Z(\vec{p}^2)}{d(\vec{p}^2)^2}. \tag{35}$$

In addition, however, as was pointed out in the last section, discretisation errors could be significant for heavy-light mesons, and we saw how the changes suggested in eqs.(18-20) lead to a better understanding of the measured data. By including these corrections and using the approximation of eq. (33), $R_2^{(2)}(t)$ for extended interpolating operators can be approximated by :

$$R_2^{(2)}(t) \approx -\frac{1}{2} \left(\frac{a^2}{4\sinh^2(\frac{ma}{2})} + \frac{at}{\sinh(ma)} - 4R_Z^{(2)}\right) \times \left(1 - \frac{L}{b(t)\sinh(\frac{L}{b(t)})}\right)$$
$$- \frac{a^2}{8\cosh^2(\frac{ma}{2})} - \frac{a^2}{24} \left(\frac{2\cosh^2(\frac{ma}{2}) - 3}{\cosh^2(\frac{ma}{2})}\right) \times \frac{L}{c(t)\sinh(\frac{L}{c(t)})}, \tag{36}$$

$$b(t) = \sqrt{\frac{b_1(t)}{b_2(t)}}, \tag{37}$$

$$b_1(t) = \frac{3a^4}{16\sinh^4(\frac{ma}{2})} + \frac{3a^3 t}{4\sinh^2(\frac{ma}{2})\sinh(ma)} + \frac{t^2 a^2}{\sinh^2(ma)}$$
$$- 8R_Z^{(2)} \left(\frac{a^2}{4\sinh^2(\frac{ma}{2})} + \frac{at}{\sinh(ma)}\right) + 8((R_Z^{(2)})^2 + 2R_Z^{(4)}), \tag{38}$$

$$b_2(t) = \frac{a^2}{4\sinh^2(\frac{ma}{2})} + \frac{at}{\sinh(ma)} - 4R_Z^{(2)}, \tag{39}$$

$$c(t) = \sqrt{c_f(t) - 8R_Z^{(2)}}, \qquad (40)$$

$$c_f(t) = \frac{a^2}{60 - 40\cosh^2(\frac{ma}{2})} \times \left(8\cosh^2(\frac{ma}{2}) + \frac{45}{\cosh^2(\frac{ma}{2})} + \frac{30}{\sinh^2(\frac{ma}{2})} + \frac{180t}{a\sinh(ma)}\right), \qquad (41)$$

where we have neglected any contributions from excited states.

The corresponding expression when the continuum energy-momentum dispersion relation is used, can be found by taking the limit $a \to 0$ in eq. (36).

Unlike the approximation used in deriving eq. (16), in obtaining eq. (36) the lattice spacing has been kept finite. We have performed exactly the sum over $x_3$ in eq. (8) and in calculating the sum over $n$, we have expanded the terms of the series in powers of $\pi n/L$ to second order (which requires an expansion of $Z(n^2 p_{min}^2)$ up to fourth order), and we have extended the limits of the sum between $-\infty$ to $+\infty$. The accuracy of the different approximations involved in the derivation of eq. (36) can only be rigorously checked in the case of local interpolating operators. In that case, $R_2^{(2)}(t)$ can be computed exactly and can be compared with the approximation of eq. (36) setting $R_Z^{(2),(4)}$ to zero. The results of this comparison are quite similar to those presented previously in Figs.1 and 2, where possible finite-lattice-spacing effects were ignored[14]. More specifically, for the physical situation which will be studied below (range of masses, $ma = 0.6 - 0.85$, range of times, $t = 11 - 16$ and for a spatial volume $24^3$) we have found that eq. (36) for local interpolating operators introduces an error of the order of 1.5% with respect to the exact result, computed by using eq. (14) and eqs.(18 - 20).

In the case of extended interpolating operators, the validity of eq. (36) depends on the accuracy of the expansion of eq. (33), and will have to be checked by analysing the measured data. The numerical values of $R_Z^{(4)}$ and $R_Z^{(2)}$ will of course depend on the specific details of the smearing procedure. In this study we use gauge-invariant Jacobi smearing on the heavy-quark field, described in detail in ref. [19].

In principle, our preferred strategy is to constrain the mass of the meson to the value obtained from the behaviour of the heavy-light two-point function $C_2^{(0)}(t)$ and to perform a two-parameter fit to the ratio $R_2^{(2)}(t)$ in order to obtain $R_Z^{(2)}$ and $R_Z^{(4)}$. However, for the expected range of values of $R_Z^{(4)}$, i.e. $R_Z^{(4)} = O(R_Z^{(2)} a^2)$, $R_2^{(2)}(t)$ is almost independent of $R_Z^{(4)}$ ($R_Z^{(4)}$ only appears in the volume correction term and its contribution is suppressed by powers of $1/t$). Thus in practice, two-parameter fits prove to be very unstable; relatively small changes in $R_Z^{(2)}$ can be compensated for by large ones in $R_Z^{(4)}$. Because of the instability of the fits with two parameters, we modify our procedure as follows: we take as our central value $R_Z^{(4)} = 0$ and for $R_Z^{(2)}$ we take the value obtained from a single parameter fit to $R_2^{(2)}$. To estimate the error in $R_Z^{(2)}$ we repeat the fits with other values of $R_Z^{(4)}$ as explained in the following paragraphs. A comparison of the predictions using this procedure with the measured data will serve as a further check of the accuracy of the several approximations performed.

---

[14]Eq.(36) for local interpolating operators and in the limit $a \to 0$ reduces to eq. (16).

|  | $R_Z^{(2)}a^{-2}$ | $R_Z^{(4)}a^{-4}$ |
|---|---|---|
| $\kappa_h = 0.125$ | $-2.75\ ^{+22}_{-20}$ | $2.1\ ^{+1.4}_{-1.7}$ |
| $\kappa_h = 0.133$ | $-2.38\ ^{+21}_{-20}$ | $0.0\ ^{+2.3}_{-2.9}$ |

Table 1: Results for $R_Z^{(2),(4)}$ extracted from ordinary two-point functions fits.

From the analysis of the usual heavy-light two-point functions ($C_2^{(0)}(t)$) we know the wave-function factors $Z(0), Z(p^2 = p_{min}^2)$ and $Z(p^2 = 2p_{min}^2)$ and therefore we can estimate the size of $R_Z^{(2),(4)}$. In Table 1 we present the values for the first and second derivatives of the wavefunction factor $Z$ obtained in this way. The results of Table 1, although affected by discretisation errors, show that in the conditions of the present study, the ratio $R_Z^{(4)}a^{-2}/R_Z^{(2)}$ is indeed of order one for both heavy-quark masses considered, and one would expect a negligible dependence of $R_2^{(2)}(t)$ on the exact value taken for $R_Z^{(4)}$.

In Table 2 we present the results for $R_Z^{(2)}$ from different fits (including or neglecting correlations between different time slices) to $R_2^{(2)}(t)$ using eq. (36), in all of which $R_Z^{(4)}$ has been fixed to zero[15]. The meson masses have been constrained to the values obtained from measuring the usual heavy-light two-point functions. We see that the values obtained for $R_Z^{(2)}$ agree well with those quoted in Table 1. Moreover all the fits presented in Table 2 have small values of $\chi^2$/dof, all of which supports the approximations used to account for the finite-volume corrections. In Figure 6, we show the measured values together with the fits using eq. (36) with $R_Z^{(4)} = 0$, for both values of $\kappa_h$. As can be seen in these plots, the agreement between data and our predictions is good and gives us confidence in our procedure.

In the next subsection, we will use the values obtained for $R_Z^{(2)}$ here as input into our calculation of the slope of the Isgur-Wise function at zero-recoil. Fixing $R_Z^{(4)}$ to zero will introduce a small systematic error in our value of the slope, which will have to be estimated. Our strategy will be the following: i) For each $\kappa_h$, the central value of the slope will be calculated by using the values of $R_Z^{(2)}$ obtained with the method outlined in this subsection

---

[15]For simplicity in the discussion, we do not present results obtained by using the continuum energy-momentum dispersion relation. In general, results obtained in this way reveal pronounced discrepancies between $R_2^{(2)}(t)$ and $C_2^{(0)}(t)$, and have considerably higher values of $\chi^2$/dof than those quoted in Table 2.

|  | $\kappa_h = 0.125, \quad R_Z^{(4)} a^{-4} = 0$ | | $\kappa_h = 0.133, \quad R_Z^{(4)} a^{-4} = 0$ | |
|---|---|---|---|---|
|  | $R_Z^{(2)} a^{-2}$ | $ma$ | $R_Z^{(2)} a^{-2}$ | $ma$ |
| corr. | $-2.90 \, ^{+5}_{-5}$ | $[0.822 \, ^{+4}_{-2}]$ | $-2.51 \, ^{+5}_{-7}$ | $[0.598 \, ^{+3}_{-1}]$ |
| $\chi^2/\text{dof}$ | 1.8/5 | . | 5.5/5 | . |
| uncorr. | $-2.91 \, ^{+5}_{-5}$ | $[0.822 \, ^{+4}_{-2}]$ | $-2.47 \, ^{+6}_{-6}$ | $[0.598 \, ^{+3}_{-1}]$ |
| $\chi^2/\text{dof}$ | 0.3/5 | . | 2.8/5 | . |

Table 2: Results for $R_Z^{(2)}$ for different fitting procedures. The fits were done for timeslices $11, \ldots, 16$. We present results obtained by using both correlated and uncorrelated fits. The masses have been obtained from $C_2^{(0)}(t)$, and are shown in square brackets.

(i.e. $R_Z^{(4)} = 0$). ii) In each case, we will recompute $R_Z^{(2)}$ and the slope of the Isgur-Wise function with two new fixed values of $R_Z^{(4)}$ given by three standard deviations from the central values of $R_Z^{(4)}$ quoted in Table 1. The spread of values obtained in this way will enable us to estimate the systematic uncertainties in our result for the slope associated with the approximations discussed in this subsection.

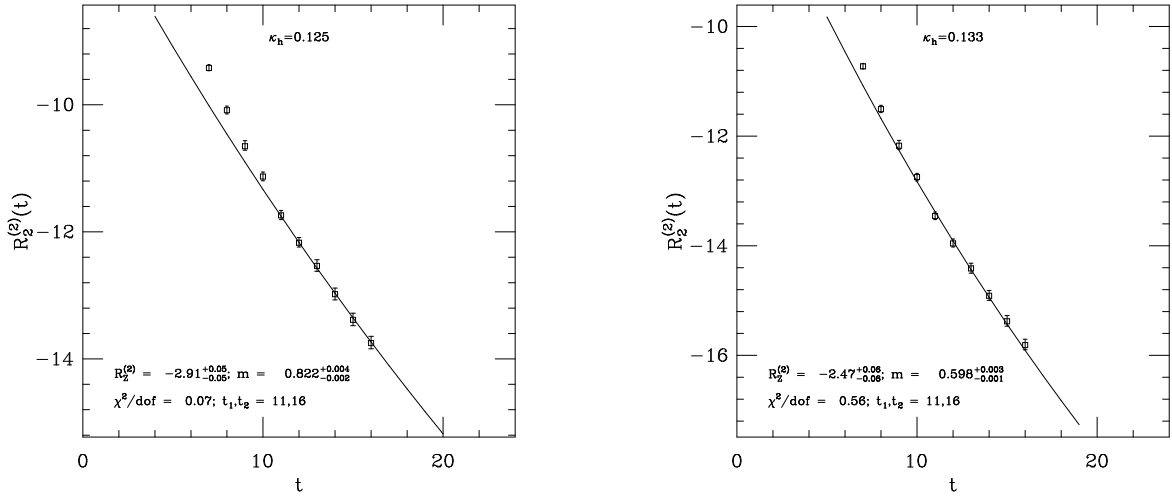

Figure 6: Ratios $R_2^{(2)}(t)$ for $\kappa_h = 0.125$ and $\kappa_h = 0.133$, together with the (uncorrelated) fits with masses constrained to their known values from ordinary two-point correlators.

## 3.2 Extended interpolating operators : The ratio $R_3^{(2)}(t)$.

The exact computation of $R_3^{(2)}(t)$, eq. (22), would require the knowledge of the wavefunction factors $Z(n^2 p_{min}^2)$ as well as the matrix elements $M^4(n^2 p_{min}^2)$ for all the possible momenta of the lattice. Following the ideas of the last subsection, we will expand $Z(n^2 p_{min}^2)$ in powers of $p_{min}$ (eq. (33)), and similarly we will use an expansion for $M^4(n^2 p_{min}^2)$ of the type:

$$M^4(n^2 p_{min}^2) \approx M^4(0) \left(1 + (\frac{2\pi}{L})^2 n^2 R_M^{(2)} + (\frac{2\pi}{L})^4 n^4 R_M^{(4)}\right), \tag{42}$$

$$R_M^{(4)} = \frac{1}{2} \frac{M^{4''}(0)}{M(0)}, \tag{43}$$

$$M^{4''}(\vec{p}^2) = \frac{d^2 M(\vec{p}^2)}{d(\vec{p}^2)^2}, \tag{44}$$

with $R_M^{(2)}$ defined in eq. (29).

With the expansions of eqs. (33) and (42) and using the same kind of approximations as in the case of $R_2^{(2)}$, $R_3^{(2)}$ can be approximated by[16]

$$R_3^{(2)}(t) \approx -\frac{1}{2}\left(\frac{a^2}{4\sinh^2(\frac{ma}{2})} + \frac{at}{\sinh(ma)} - 2(R_Z^{(2)} + R_M^{(2)})\right) \times \left(1 - \frac{L}{b'(t)\sinh(\frac{L}{b'(t)})}\right)$$

$$-\frac{a^2}{8\cosh^2(\frac{ma}{2})} - \frac{a^2}{24}\left(\frac{2\cosh^2(\frac{ma}{2}) - 3}{\cosh^2(\frac{ma}{2})}\right) \times \frac{L}{c'(t)\sinh(\frac{L}{c'(t)})}, \tag{45}$$

where $b'(t)$ and $c'(t)$ are given by the expressions for $b(t)$ and $c(t)$ of eqs.(37-41) with the following substitutions:

$$((R_Z^{(2)})^2 + 2R_Z^{(4)}) \rightarrow (R_Z^{(2)} R_M^{(2)} + (R_Z^{(4)} + R_M^{(4)})), \tag{46}$$

$$2R_Z^{(2)} \rightarrow R_Z^{(2)} + R_M^{(2)}. \tag{47}$$

In order to compute eq. (45), we set $R_Z^{(4)}$ to zero and we use the values for $R_Z^{(2)}$ extracted from the fits to the ratio $R_2^{(2)}(t)$, as input. Thus, we have two free parameters, $R_M^{(2)}$ and $R_M^{(4)}$, (involving the first and the second derivatives of the Isgur-Wise function at the zero-recoil point) which, in principle, should be determined from the fit to the ratio $R_3^{(2)}(t)$. $R_M^{(2)}$ contains $\xi'(1)$, eq. (29), whereas $R_M^{(4)}$ is given by:

$$R_M^{(4)} = \frac{1}{8m^4}(\frac{d^2\bar{\xi}}{d\omega^2}(1) - \frac{1}{2}). \tag{48}$$

---

[16]In eq. (45) the "lattice" energy-momentum dispersion relation of eqs.(18- 20) has been used. The expression for the ordinary continuum energy-momentum dispersion relation could be found by taking the limit $a \to 0$ in eq. (45).

The derivatives of the radiative corrections, which determine the relation between the derivatives of $\xi$ and $\bar{\xi}$ at $\omega = 1$, were calculated in [7] and [8], and around the mass of the charm quark (our $\kappa_h = 0.125$ ($\kappa_h = 0.133$) quark is slightly heavier (lighter) than the charm quark, [16]) are approximately equal to $\beta'(1) = -0.24$ and $\beta''(1) = 0.17$.

As in the case of $R_2^{(2)}(t)$ and $R_Z^{(4)}$, for those values of $R_M^{(2)}$ and $R_M^{(4)}$ for which eq. (42) is a valid expansion, $R_3^{(2)}(t)$ hardly depends on $R_M^{(4)}$ and in the limit $L \to \infty$ is totally independent of it. On the other hand, we can get an estimate of the size of the ratio $(R_M^{(4)} \times m^2/R_M^{(2)})$ by assuming a theoretical dependence of the Isgur-Wise function on $\omega$, in a region close to $\omega = 1$. We have studied two different dependencies, widely used in the literature [20]:

$$\xi(\omega) = \frac{2}{\omega + 1} \exp\left(-(2\rho^2 - 1)\frac{\omega - 1}{\omega + 1}\right), \tag{49}$$

$$\xi(\omega) = \exp\left(-\rho^2(\omega - 1)\right). \tag{50}$$

In both cases, for values of $\rho^2 \approx 1.5$ [5], we find that the ratio $(R_M^{(4)} \times m^2/R_M^{(2)})$ takes values of the order of 0.5. Therefore, the contribution of the $R_M^{(4)}$ term to the expansion of eq. (42) is suppressed by at least an extra factor $p_{min}^2$ with respect to the $R_M^{(2)}$ contribution, and furthermore the $R_M^{(4)}$ contribution to $R_3^{(2)}(t)$ will be also suppressed by powers of $1/t$. Because of the instability of the fits with two parameters we have followed two procedures:

i) By using one of the $\omega$ dependencies of eqs.(49-50), we can rewrite $R_M^{(4)}$ in terms of $R_M^{(2)}$ and therefore perform just a one-parameter fit to the ratio $R_3^{(2)}(t)$ in order to obtain $R_M^{(2)}$. We have checked that the difference in the final result for the slope of the Isgur-Wise function at $\omega = 1$ depends only at the level of 1% on the particular $\omega$ dependence used.

ii) Another possibility is to fix $R_M^{(4)}$ to zero; in this way $R_M^{(2)}$ can again be extracted from a one-parameter fit to $R_3^{(2)}(t)$. The results obtained in this way differ from the previous ones only at the level of 1-1.5% for $\kappa_h = 0.125$ and at the level of 3% for $\kappa_h = 0.133$.

To obtain our final results, we have decided to use method (ii). In this way, our estimate for the slope of the Isgur-Wise function will not depend on any particular choice of the theoretical dependence of the Isgur-Wise function on $\omega$. The systematic error associated with this approximation, as we have discussed, could be of the order of 1.5% for $\kappa_h = 0.125$ and 3% for $\kappa_h = 0.133$, which is smaller than other sources of errors present in our calculation.

In Table 3, we present our results for $\bar{\xi}'(1)$. We quote numbers for $\bar{\xi}'(1)$ obtained by fitting $R_3^{(2)}(t)$ to the functional form of eq. (45) and by fitting the difference of ratios $R_3^{(2)}(t) - R_2^{(2)}(t)$ to the difference between the functions of eq. (45) and eq. (36). In all of the cases, $R_Z^{(2)}$ and $m$ have been obtained from our previous study of $R_2^{(2)}(t)$ and $C_2^{(2)}(t)$, respectively. The best $\chi^2/\text{dof}$ are obtained when we look at the difference of ratios $R_3^{(2)}(t) - R_2^{(2)}(t)$. Presumably, in this difference of ratios, the finite-volume corrections partially cancel and, therefore, results are less sensitive to the exact details of the procedure.

| $\kappa_h = 0.125$ | $R_3^{(2)}$ | $R_3^{(2)} - R_2^{(2)}$ | $\kappa_h = 0.133$ | $R_3^{(2)}$ | $R_3^{(2)} - R_2^{(2)}$ |
|---|---|---|---|---|---|
| | $R_Z^{(2)} a^{-2}$ | $\bar{\xi}'(1)$ | $\bar{\xi}'(1)$ | $R_Z^{(2)} a^{-2}$ | $\bar{\xi}'(1)$ | $\bar{\xi}'(1)$ |
| corr. | $-2.90\,^{+5}_{-5}$ | $-2.00\,^{+16}_{-15}$ | $-1.91\,^{+14}_{-13}$ | $-2.51\,^{+5}_{-7}$ | $-1.75\,^{+9}_{-13}$ | $-1.61\,^{+9}_{-7}$ |
| $\chi^2/\text{dof}$ | 1.8/5 | 5.0/2 | 0.7/2 | 5.5/5 | 15.9/2 | 0.2/2 |
| uncorr. | $-2.91\,^{+5}_{-5}$ | $-1.92\,^{+16}_{-14}$ | $-1.94\,^{+16}_{-15}$ | $-2.47\,^{+6}_{-6}$ | $-1.63\,^{+8}_{-7}$ | $-1.59\,^{+10}_{-8}$ |
| $\chi^2/\text{dof}$ | 0.3/5 | 0.0/2 | 0.1/2 | 2.8/5 | 0.4/2 | 0.0/2 |

Table 3: Results for $\bar{\xi}'(1)$ for different fitting procedures. The fits to the ratio of two-point correlators have been extracted from Table 2. The ratio of three-point correlators was fitted for timeslices $11, 12, 13$, (in our simulation $t_y$ in eq. (22) is equal to 24).

As can be seen in Table 3 the results are rather stable under variations of the method used. For $\kappa_h = 0.125$ a reasonable value for $\bar{\xi}'(1)$ could be $-1.92\,^{+16}_{-14}$. This is the value obtained using an uncorrelated fit with the mass of the meson constrained to its known value from ordinary two-point function fits. Correlated fits in principle give a more realistic estimate of the quality of the fit as measured by $\chi^2$; however they are more sensitive to the uncertainties in the fitting function. The errors quoted for $\bar{\xi}'(1)$ are only statistical and to them must be added those associated with our treatment of $R_Z^{(4)}$ and $R_M^{(4)}$. As was discussed at the end of the last subsection, in order to estimate the uncertainties due to our ignorance of $R_Z^{(4)}$, we have recomputed $\bar{\xi}'(1)$ for two more fixed values of $R_Z^{(4)}$ ($R_Z^{(4)} = -3$ and $R_Z^{(4)} = 6$) obtained by allowing three standard deviations from the central value of $R_Z^{(4)}$ quoted in Table 1. The spread of values obtained in this way for $\bar{\xi}'(1)$ determines a further error of around of 1%-2%. This error, together with the 1.5% discussed before due to $R_M^{(4)}$, gives a final estimate of our systematic uncertainties of the order of 3%. Thus, the value for $\bar{\xi}'(1)$ quoted above, together with our estimate for the systematic uncertainties, leads to a result for the slope of the Isgur-Wise function at the zero-recoil point, for $\kappa_h = 0.125$, of

$$\xi'(1) = \bar{\xi}'(1) - \beta'(1) = -1.7\,^{+2}_{-2} \qquad (51)$$

where we have taken $\beta'(1) = -0.24$ ([7]).

For $\kappa_h = 0.133$, there is a greater uncertainty. In this case a correlated fit to $R_3^{(2)}$ has a large $\chi^2/\text{dof}$, which may be an indication of a poorly-known fitting function. However, a correlated fit to the difference between ratios, $R_3^{(2)}(t) - R_2^{(2)}(t)$, gives a good $\chi^2/\text{dof}$ and an answer for $\bar{\xi}'(1)$ in excellent agreement with the answer from uncorrelated fits. Thus, following the same criterion as for $\kappa_h = 0.125$, we quote a value for $\bar{\xi}'(1) = (-1.63)\,^{+8}_{-7}$ or, equivalently, a value for the slope of the Isgur-Wise function at the zero-recoil point of

$$\xi'(1) = \bar{\xi}'(1) - \beta'(1) = -1.4\,^{+2}_{-1} \qquad (52)$$

where we have included in a similar way to that above, the systematic errors associated with our choice of $R_Z^{(4)}$ and $R_M^{(4)}$, and, for this value of the quark mass, we have taken $\beta'(1) = -0.20$ ([7]).

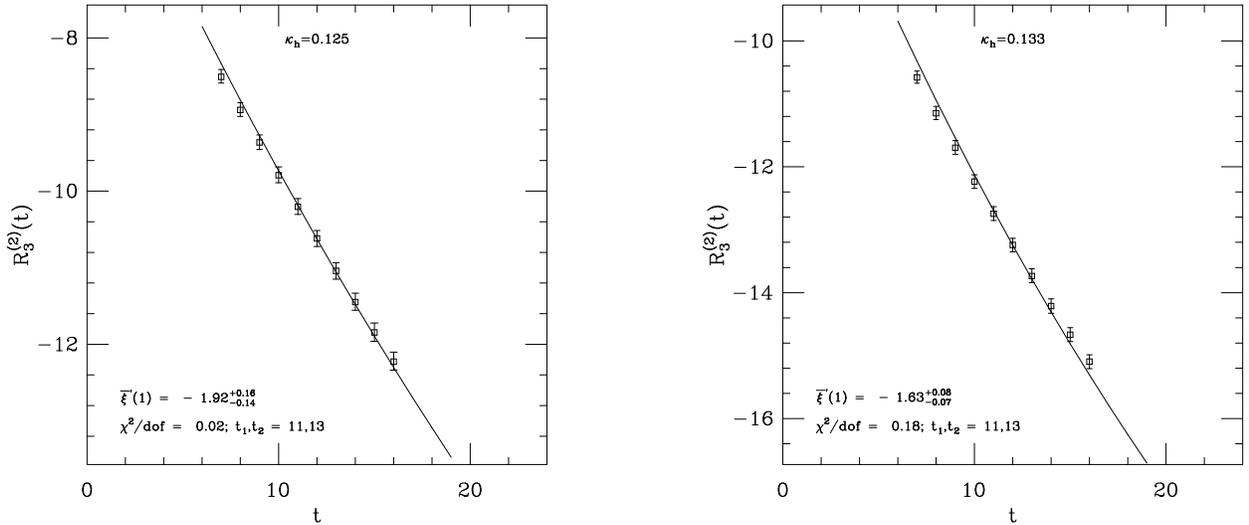

Figure 7: Ratios $R_3^{(2)}(t)$ for $\kappa_h = 0.125$ and $\kappa_h = 0.133$, together with the (uncorrelated) fits with the masses constrained to their known values from ordinary two-point correlators.

The values of eqs.(51-52) may be slightly higher than, but certainly compatible within errors with, the results of [7] given in the introduction, obtained with the traditional method of extrapolating from $\omega > 1$ to $\omega = 1$.

In Figure 7 we show the ratio $R_3^{(2)}(t)$ together with the fit, for both values of $\kappa_h$.

By construction, the Isgur-Wise function, and therefore its derivatives, are independent of the mass of the heavy quark considered. Although within errors our results for the two heavy quarks studied here agree, the central values (eqs. (51-52)) may suggest some dependence on the heavy-quark mass. Trying to conclude that this effect is not merely statistical and trying to quantify the size of the $O(1/m_Q)$ corrections (derivatives at the zero-recoil point of $\gamma(\omega)$, defined in eq. (24)) would require an extensive study, involving more than two heavy quarks, which is outside the exploratory aim of this paper. Such a study is made, using the traditional method of extrapolating from $\omega > 1$ to $\omega = 1$, in ref. [7].

Recently the CLEO collaboration has reported a value for $\xi'(1) = 1.01\pm0.15\pm0.09$ obtained by looking at the semileptonic decay $B \to D^*$, [21]. This number should not be compared directly with our results (eqs. (51-52)), as our results were obtained using a light antiquark heavier than the strange quark. However sum-rules ([22]) and quark-model ([23], [24]) calculations suggest that the absolute value of the slope of the Isgur-Wise function at zero-recoil decreases with the mass of the light antiquark.

## 4 Conclusions

In this paper we have presented the results from an exploratory study implementing the method suggested in ref.[8] for computing the slope of the Isgur-Wise function at the zero-recoil point in lattice simulations. We have found large finite-volume corrections on currently available lattices, effects which will be present in all direct computations of derivatives of matrix elements with respect to external momenta. We have studied these corrections and have given their volume, mass and time dependence. We have seen that these finite-volume corrections do not have a dynamical origin, in contrast to those studied in [10]-[12], and the usual two- and three-point correlation functions (without the $x_3^2$ insertion) are not affected by them. Eq.(17) constitutes a simple way of estimating the expected size of these corrections for a given simulation. Furthermore, from this study, we have learnt how to account, in an approximate but sufficiently precise way, for these corrections and have been able to obtain reasonable values ($\xi'(1) = -1.7\ ^{+2}_{-2}$ and $\xi'(1) = -1.4\ ^{+2}_{-1}$) for the slope of the Isgur-Wise function, for two mesons composed of a heavy quark slightly heavier ($\kappa_h = 0.125$) and lighter ($\kappa_h = 0.133$), respectively, than the charm quark, and in both cases, a light antiquark whose mass is about that of the strange quark ($\kappa_l = 0.14144$). These values compare well with those obtained in [7]. It would be interesting to study the dependence of the results on both heavy- and light-quark masses.

We have seen that the understanding of the finite-volume effects is considerably simpler; indeed they seem to be fully under control, if local interpolating operators are used. The price for using local operators is that their overlap with the ground state is generally poorer, and one has to fit correlation functions at larger times in order to be able to neglect the contributions from excited states. Nevertheless, the calculation described in this paper should be repeated with local interpolating operators.

### Acknowledgements

We thank K.C. Bowler for a critical reading of the manuscript. This work has been partially supported by the European Union, under contract No. CHRX-CT92-0051. CTS and DGR acknowledges the Particle Physics and Astronomy Research Council for their support through the award of a Senior Fellowship and Advanced Fellowship respectively. JN acknowledges the European Union for for their support through the award of a Postdoctoral Fellowship, contract No. CHBICT920066. NS thanks the Noopolis-Soven Foundation for financial support.

# References

[1] M. Neubert, SLAC preprint, SLAC-PUB-6263 (1993).

[2] ARGUS Collaboration, H. Albrecht et al., Z. Phys. C57 (1993) 533.

[3] CLEO Collaboration, G. Crawford et al., " A Measurement of $B(\bar{B}^0 \to D^{*+}l^-\bar{\nu})$", contributed paper to the 1993 Photon-Lepton Conference, and references there in.

[4] N. Isgur and M.B. Wise, Phys. Lett. B232 (1989) 113; Phys. Lett. B237 (1990) 527.

[5] UKQCD Collaboration, S.P. Booth et.al., Phys. Rev. Lett. 72 (1994) 462.

[6] C.W. Bernard, Y. Shen and A. Soni, Phys. Lett. B317 (1993) 164; Nucl. Phys. B. (Proc. Suppl.)30 (1993) 473.

[7] UKQCD collaboration, "Semi-Leptonic Heavy-Light→Heavy-Light Meson Decays", presented by L. Lellouch at The XII International Symposium on Lattice Field Theory, Bielefeld (1994). UKQCD Collaboration, in preparation.

[8] U. Aglietti, G. Martinelli and C.T. Sachrajda, Phys. Lett. B 324 (1994) 85.

[9] J.E.Mandula and M.C.Ogilvie, Nucl. Phys. B (Proc. Suppl.) 34 (1994) 480

[10] M. Lüscher, Comm. Math. Phys. 104 (1986) 177, and references therein.

[11] S. Aoki et. al., Nucl. Phys. B (Proc. Suppl.) 34 (1994) 363.

[12] M.Fukugita et. al., Phys. Lett. B294 (1992) 350.

[13] UKQCD Collaboration, C. Allton et.al., Phys. Rev. D49, (1994) 474.

[14] B. Sheikholeslami and R. Wohlert, Nucl. Phys. B259 (1985) 572.

[15] G. Heatlie, C.T. Sachrajda, G. Martinelli, C. Pittori and G.C. Rossi, Nucl. Phys. B352 (1991) 266.

[16] UKQCD Collaboration, R.M. Baxter et.al., Phys. Rev. D49, (1994) 1594.

[17] G.Martinelli, S.Petrarca, C.T.Sachrajda and A.Vladikas, Phys. Lett. B311 (1993) 241

[18] M. Creutz, "Quarks, gluons and lattices",(Cambridge Monographs on Mathematical Physics,1983) p.1.

[19] UKQCD Collaboration, C. Allton et.al., Phys. Rev. D47, (1993) 5128.

[20] M. Neubert, V. Rieckert, B. Stech, and Q.P. Xu, in "Heavy Flavours", edited by A.J. Buras and M. Lindner (World Scientific, Singapore)

[21] CLEO collaboration, "Measurement of $|V_{cb}|$ from $B \to D^* l \bar{\nu}$ decays at the T(4s)", presented by T. Browder at the 27th International Conference on High Energy Physics, Glasgow (1994) .

[22] T. Huang, "Light quark dependence of the Isgur-Wise function", presented at the 27th International Conference on High Energy Physics, Glasgow (1994). T. Huang and C.W. Luo, IHEP Sinica preprint BIHEP-TH-94-10.

[23] F.E. Close, "Flavour dependence of form factors in heavy meson decays", presented at the 27th International Conference on High Energy Physics, Glasgow (1994).

[24] M. Sadzikowsky and K. Zalewski, Z. Phys. C59 (1993), 677.